\documentclass[pdflatex,sn-mathphys-num]{sn-jnl}

\usepackage{graphicx}%
\usepackage{multirow}%
\usepackage{amsmath,amssymb,amsfonts}%
\usepackage{amsthm}%
\usepackage{mathrsfs}%
\usepackage[title]{appendix}%
\usepackage{xcolor}%
\usepackage{textcomp}%
\usepackage{manyfoot}%
\usepackage{booktabs}%
\usepackage{algorithm}%
\usepackage{algorithmicx}%
\usepackage{algpseudocode}%
\usepackage{listings}%
\theoremstyle{thmstyleone}%
\theoremstyle{thmstyletwo}%

\theoremstyle{thmstylethree}%

\begin{document}

\title[Article Title]{Spontaneously formed phonon frequency combs in van der Waals solid CrXTe$_3$ (X=Ge,Si)}


\author*[1,2]{\fnm{Lebing} \sur{Chen}}\email{lebingchen@berkeley.edu}
\equalcont{These authors contributed equally to this work.}

\author[3]{\fnm{Gaihua} \sur{Ye}}%\email{}
\equalcont{These authors contributed equally to this work.}

\author[3]{\fnm{Cynthia} \sur{Nnokwe}}%\email{}

\author[4]{\fnm{Xing-Chen} \sur{Pan}}%\email{}

\author[4,5]{\fnm{Katsumi} \sur{Tanigaki}}%\email{}

\author[4,6,7]{\fnm{Guanghui} \sur{Cheng}}

\author[6,7,4,8]{\fnm{Yong P.} \sur{Chen}}

\author[9,10]{\fnm{Jiaqiang} \sur{Yan}}

\author[9,10]{\fnm{David G.} \sur{Mandrus}}

\author[6]{\fnm{Andres E.} \sur{Llacsahuanga Allcca}} 

\author[1,2]{\fnm{Nathan} \sur{Giles-Donovan}}

\author*[1,2]{\fnm{Robert J.} \sur{Birgeneau}}\email{robertjb@berkeley.edu}

\author*[3]{\fnm{Rui} \sur{He}}\email{rui.he@ttu.edu}

\affil*[1]{\orgdiv{Department of Physics}, \orgname{University of California}, \orgaddress{\city{Berkeley}, \state{California}, \postcode{94720},  \country{USA}}}

\affil[2]{\orgdiv{Material Sciences Division}, \orgname{Lawrence Berkeley National Laboratory}, \orgaddress{\city{Berkeley}, \state{California}, \postcode{94720}, \country{USA}}}

\affil[3]{\orgdiv{Department of Electrical and Computer Engineering}, \orgname{Texas Tech University}, \orgaddress{\city{Lubbock}, \state{Texas}, \postcode{79409}, \country{USA}}}

\affil[4]{\orgdiv{WPI Advanced Institute for Materials Research (AIMR)}, \orgname{Tohoku University}, \orgaddress{\city{Sendai}, \postcode{980-8577}, \country{Japan}}}

\affil[5]{\orgdiv{Department of Physics, Graduate School of Science}, \orgname{Tohoku University}, \orgaddress{\city{Sendai}, \postcode{980-8578}, \country{Japan}}}

\affil[6]{\orgdiv{Department of Physics and Astronomy}, \orgname{Purdue University}, \orgaddress{\city{West Lafayette}, \state{Indiana}, \postcode{47907}, \country{USA}}}

\affil[7]{\orgdiv{School of Electrical and Computer
Engineering and Purdue Quantum Science and Engineering Institute}, \orgname{Purdue University}, \orgaddress{\city{West Lafayette}, \state{Indiana}, \postcode{47907}, \country{USA}}}

\affil[8]{\orgdiv{Institute of Physics and Astronomy and Villum Centers for Dirac Materials and for Hybrid Quantum Materials}, \orgname{Aarhus University}, \orgaddress{\city{Aarhus-C}, \country{Denmark}}}

\affil[9]{\orgdiv{Department of Materials Science and Engineering}, \orgname{University of Tennessee}, \orgaddress{\city{Knoxville}, \state{Tennessee}, \postcode{37996}, \country{USA}}}

\affil[10]{\orgdiv{Materials Science and Technology Division}, \orgname{Oak Ridge National Laboratory}, \orgaddress{\city{Oak Ridge}, \state{Tennessee}, \postcode{37831}, \country{USA}}}


\abstract{Optical phonon engineering through nonlinear effects has been utilized in ultrafast control of material properties. However, nonlinear optical phonons typically exhibit rapid decay due to strong mode-mode couplings, limiting their effectiveness in temperature or frequency sensitive applications. In this study, we report the observation of long-lived nonlinear optical phonons through the spontaneous formation of phonon frequency combs in the van der Waals material CrXTe$_3$ (X=Ge, Si) using high-resolution Raman scattering. Unlike conventional optical phonons, the highest $A_g$ mode in CrGeTe$_3$ splits into equidistant, sharp peaks forming a frequency comb that persists for hundreds of oscillations and survives up to 100K before decaying. These modes correspond to localized oscillations of Ge$_2$Te$_6$ clusters, isolated from Cr hexagons, behaving as independent quantum oscillators. Introducing a cubic nonlinear term to the harmonic oscillator model, we simulate the phonon time evolution and successfully replicate the observed comb structure. Similar frequency comb behavior is observed in CrSiTe$_3$, demonstrating the generalizability of this phenomenon. Our findings reveal that Raman scattering effectively probes high-frequency nonlinear phonon modes, providing new insight into generating long-lived, tunable phonon frequency combs with applications in ultrafast material control and phonon-based technologies.}

\maketitle

\section*{Introduction}\label{sec1}

Optical phonon engineering with nonlinearity is extensively used to control material properties in an ultrafast manner. For instance, ultrafast light pulses can be used to achieve rapid control over magnetism, superconductivity, and ferroelectricity\cite{Nova2019,Nova2017,Hoegen2022}. Typically, in these processes, a femtosecond light pulse is used to excite an optical phonon to a large amplitude that dynamically changes the properties of the material\cite{Disa2021}. However, nonlinear optical phonons tend to decay rapidly due to strong mode-mode couplings, especially at higher amplitudes\cite{Wu2008}. As a result, the tuning duration is often limited to fewer than 100 oscillation periods before the energy dissipates entirely as heat. This rapid decay presents challenges such as unwanted sample heating, which accelerates dissipation, and introduces measurement uncertainties in the frequency domain. Moreover, the availability of strong-field THz sources within the 5–15 THz (170-500 cm$^{-1}$) frequency range has been limited, restricting their capacity to fully cover the optical phonon spectrum for certain materials\cite{Dhillon2017,Leitenstorfer2023}. In contrast, frequency-resolved techniques such as Raman scattering and infrared spectroscopy can access phonon energies both within and beyond this range, with weaker driving amplitudes. Therefore, probing phonon nonlinearities with these techniques allow for optical phonon engineering with larger bandwidth, finer frequency resolution and greater stability against thermalization.

The decay of nonlinear optical phonons is due to anharmonic interactions in the crystal lattice, where multiple vibrational modes interact within the anharmonic atomic potential, creating numerous decay pathways that increase the likelihood of energy dissipation\cite{Barman2004}. In the frequency domain, it will be reflected as the increased linewidth of the decaying phonon mode, as observed in bulk Bi$_2$Te$_3$\cite{Buchenau2020, He2012}. Conversely, if an optical phonon is predominantly associated with a unique atomic bond and has fewer coupling pathways with other modes, it will be more difficult for this mode to decay even with anharmonicity. Instead, it may sustain parametric oscillations, leading to the formation of phonon frequency combs, while the phonon linewidth remains nearly resolution-limited. The concept of frequency comb originates from optical metrology, where it is characterized by a spectrum with uniformly spaced, narrow peaks across a range of frequencies\cite{Udem2002}. Generated from mode-locked lasers\cite{Teets1977, Eckstein1978} and Kerr resonators\cite{Kippenberg2011}, optical frequency combs are widely employed in various applications, including optical clocks\cite{Udem1999} and frequency comb spectroscopy\cite{Picqué2019}. As similar bosonic excitations, phonon frequency combs were first described in the context of the Fermi-Pasta-Ulam-Tsingou chain, where interatomic interactions include cubic or quartic terms in addition to the quadratic Hamiltonian\cite{Berman2005}. In this model, the population of states will oscillate for an extended period before eventually settling into thermal equilibrium\cite{Fermi1955}. Extensive research has been dedicated to generating phonon frequency combs using acoustic phonons in microresonators\cite{Cao2014, Ganesan2017, Ganesan2018} or Fabry-Pérot cavities\cite{Yoon2024}, where few decay paths are available. However, there are few reports on the spontaneous formation of phonon frequency combs within the optical phonon branches of crystalline solids, as most optical phonon modes typically decay. 

In this study, we present the observation of spontaneously formed frequency combs in the van der Waals solid CrXTe$_3$ (X=Ge,Si) using high-resolution Raman scattering. The Raman spectrum of CrGeTe$_3$ reveals that its highest energy $A_g$ optical phonon mode splits into several equidistant, sharp peaks which persists for hundreds of oscillations and survives up to 100K. Analysis of the phonon dispersion relations and vibrations indicates that these modes exhibit flat dispersions, involving localized oscillations of Ge$_2$Te$_6$ clusters separated by Cr hexagons. These can effectively be regarded as independent quantum oscillators. By introducing a cubic nonlinear term into the harmonic oscillator Hamiltonian as a small perturbation, we simulate the time evolution of the phonon oscillation using both analytical and numerical tools, and successfully reproduce the comb features in the frequency domain. Similar frequency combs are observed in CrSiTe$_3$ as well, which can be described by the same model but with a different set of parameters.

\section*{Raman Spectrum of CrGeTe$_3$}\label{sec2}

The overall phonon spectrum of CrGeTe$_3$ from Raman scattering is relatively well-established. CrGeTe$_3$ belongs to the rhombohedral $R\bar{3}$ space group, where hexagonal Cr layers stack with ABC-stacking configuration, with Ge$_2$Te$_6$ clusters located in the center of the Cr hexagons (Fig.1a)\cite{Carteaux1995}. The primitive rhombohedral cell contains a chemical formula of Cr$_2$Ge$_2$Te$_6$(Fig.1b), giving rise to a total of 30 phonon modes. Among these phonon modes, there are 10 Raman active modes from $E_g^1$ to $E_g^5$ and $A_g^1$ to $A_g^5$\cite{Tian2016}. %The energies of these modes are listed in Table. I with comparison from $ab$ $initio$ calculations\cite{Zhang2019, Chen2022}. 
\begin{figure*}%[t]
\centering
\includegraphics[scale=.1]{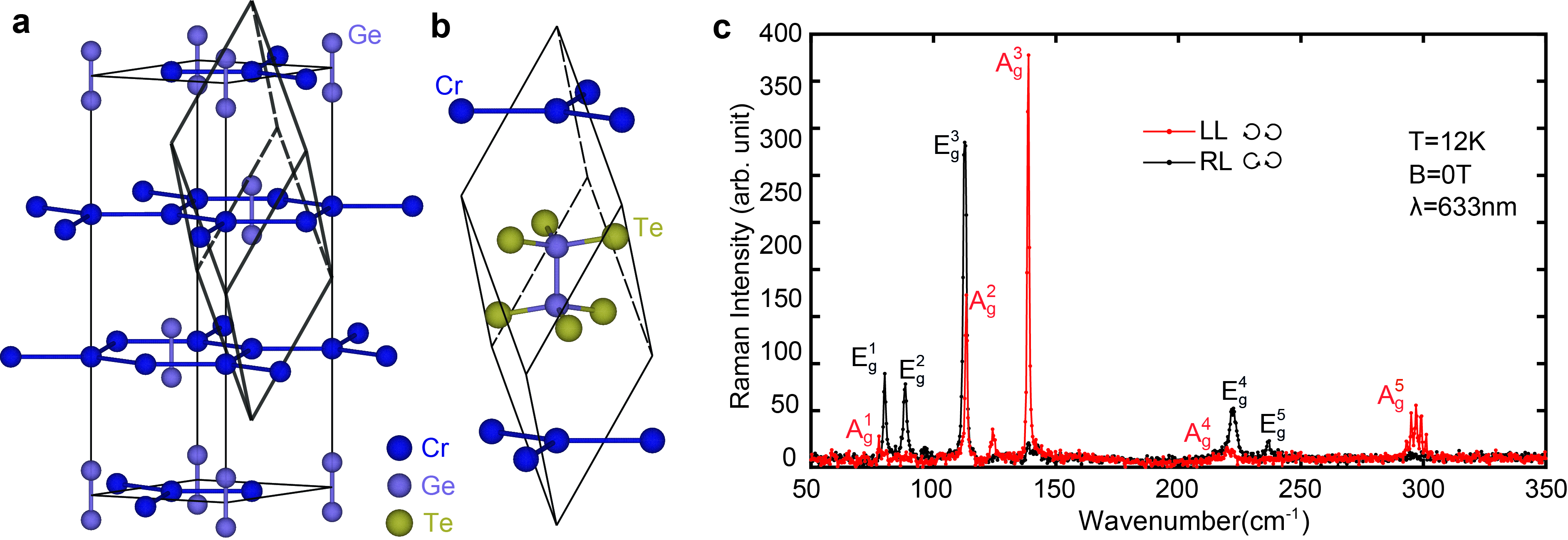}
\caption{\label{fig1} \textbf{Structure and the overall Raman spectrum of CrGrTe$_3$.}(a) Crystal structure of CrGeTe$_3$ in the hexagonal lattice unit. The Te atoms are not drawn. A rhombohedral lattice unit is drawn inside the hexagonal lattice unit. (b) The rhombohedral primitive cell of CrGeTe$_3$. (c) Overall Raman spectrum of CrGeTe$_3$ in parallel(LL) and cross (RL) circularly polarized channels.}
\end{figure*}

Fig.1c shows the Raman spectra of these modes, where the incident and scattered lights are polarized in the left-hand(L) and right-hand(R) circular polarization. All 10 Raman active modes have been observed in this spectrum, with energies consistent with previous reports and $ab$ $initio$ calculations\cite{Tian2016, Zhang2019, Chen2022}. It can be shown from Raman tensor calculations that the $E_g$ modes are only visible in the cross-polarization (LR/RL) channels and the $A_g$ modes are only visible in the parallel-polarization channels (LL/RR) (see SI)\cite{Tian2016}. 

The most prominent feature of the parallel-polarization Raman spectrum is the occurrence of frequency combs in the $A_g^5$ mode. As shown in detail in Fig.2a-2c, at low temperatures of 12K, the $A_g^5$ mode shows six sharp peaks instead of one broad peak as previously reported, with an equidistant spacing of 2cm$^{-1}$\cite{Tian2016}. %Previous reports fail to observe the frequency comb due to the resolution limit of 1.8 cm$^{-1}$, in contrast, this experiment has a much higher resolution of around 0.4cm$^{-1}$. 
To eliminate the possibility of this being a measurement artifact, we conduct a supplementary Raman scattering experiment using a 532 nm laser (Fig.2b) as a complement to the originally used 633 nm laser. These two experiments yield nearly identical outcomes, confirming the same positions and distances between the peaks. For both experiments, we fit the energy of the comb peaks $E_n$ as a function of their indices $n$: $E_n=E_0-An$, with $E_0$= 305.38$\pm$0.06cm$^{-1}$,  $A$= 1.995$\pm$0.015cm$^{-1}$ for the 633nm data, and $E_0$=305.67$\pm$0.08cm$^{-1}$, $A$=  2.017$\pm$0.021cm$^{-1}$ for the 532nm data. Both datasets show excellent linearity (fig. 2d), which is one of the defining qualities of a frequency comb. Noticeably, the linewidth of these peaks are mostly resolution-limited, with minimal broadening indicating a very small amount of phonon decay (see SI). 

\section*{Field and temperature dependence}\label{sec2}

In bulk, CrGeTe$_3$ is a ferromagnet below its Curie temperature $T_C\sim$ 65K\cite{Carteaux1995,Gong2017}, and it has been observed to exhibit significant magnon-phonon coupling\cite{Tian2016, Zhang2019,Chen2022}. This makes it crucial to investigate the potential links between the frequency combs and the spin degree of freedom, such as whether the observed splitting might be attributable to magnon scatterings in addition to phonon excitations. Two specific experiments are performed to explore the connections between magnetic order and phonon frequency combs. The first involves Raman scattering with a magnetic field applied along the $c$-axis. Figure 2e displays the field dependence of the phonon frequency comb, and there is no discernible impact on the comb's energy and spacing for magnetic fields up to 7T, suggesting that magnon excitations are not responsible for the emergence of frequency combs. This conclusion is supported by the fact that a 7T field would result in a 9.8 cm$^{-1}$ energy shift for magnons in this spin-3/2 system\cite{Li2020}. 

\begin{figure*}%[t]
\centering
\includegraphics[scale=.47]{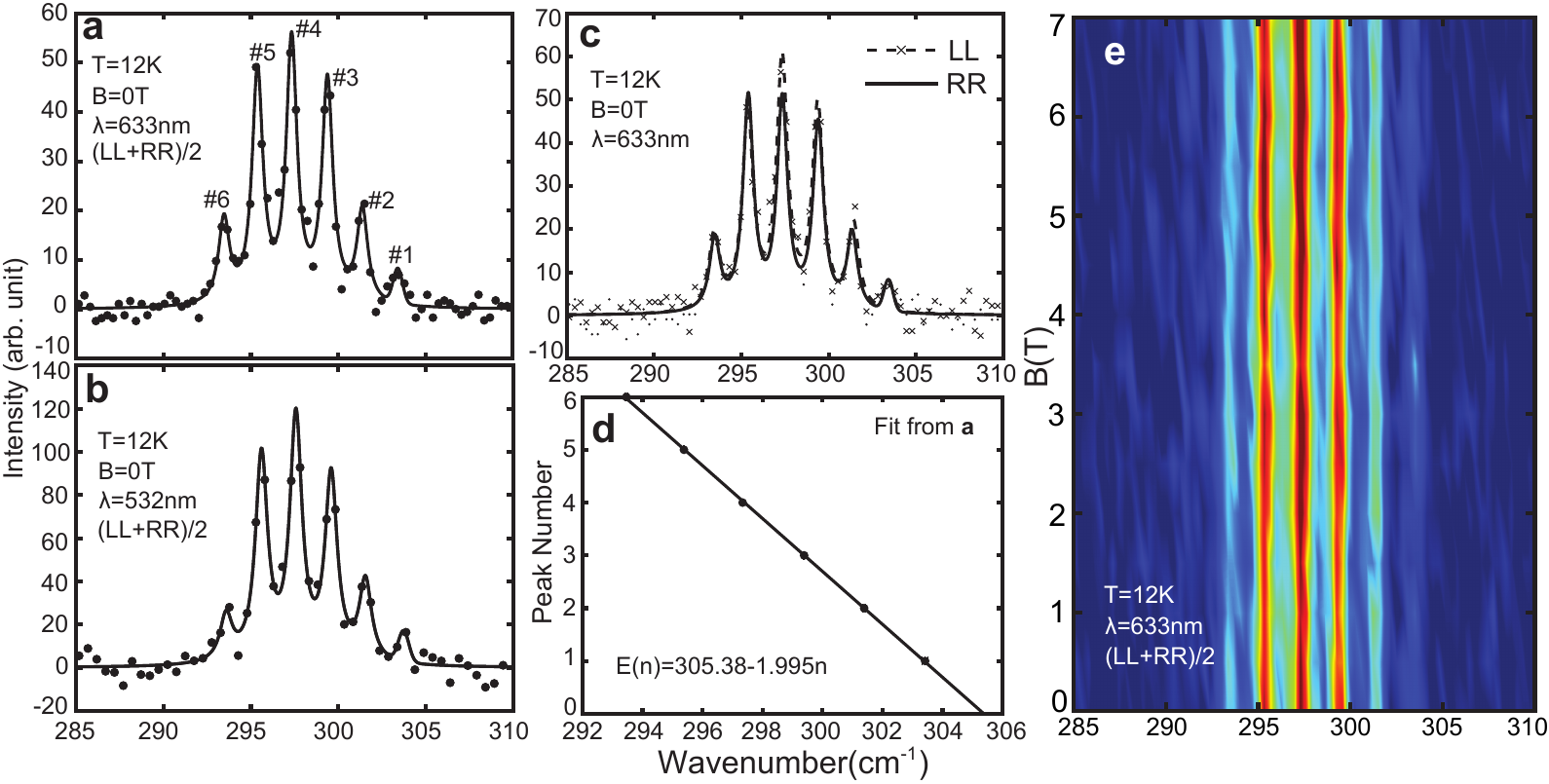}
\caption{\label{fig2} \textbf{Phonon frequency combs in CrGeTe$_3$ and its field dependence.} (a, b) Detailed views of the $A_g^5$ mode of CrGeTe$_3$ in the parallel-polarization channel, data collected with 633nm and 532nm lasers respectively. The solid lines are Lorentzian fits. (c) Comparison of frequency combs between LL and RR polarization, showing similar patterns and intensities. (d) Linear fit of the comb energy as a function of the peak index. (e) Magnetic field dependence of the frequency combs observed in the $A_g^5$ mode.}
\end{figure*}

The second experiment examines how the energies of the frequency comb vary with temperature. Should magnon-phonon coupling influence this phonon mode, the temperature dependence would deviate from the expected behavior of the standard phonon-phonon scattering model, as reported in some $E_g$ modes in CrGeTe$_3$\cite{Tian2016, Sun2018}. Fig. 3 shows the detailed temperature dependence of the $A_g^5$ mode. From the color plot in Fig. 3a and 3b, we can see that the comb feature persists above $T_C$, with the central three modes at 295.6, 297.6, and 299.6cm$^{-1}$ at 12K persisting to above 100K for the 532nm wavelength experiment (fig.3b). To verify the magnetic order's possible role in the $A_g^5$ phonon energy, we utilize the 3-phonon decay model 
\begin{equation}
E(T)=E(0)-\frac{a}{e^{\frac{E(0)}{2k_BT}}-1},
\end{equation}
where $E(0)$ is the phonon energy at zero temperature, and $a$ is related to the scattering strength of the anharmonic decay\cite{Choe2021, Buchenau2020, Klemens1966}. We fit the Raman spectrum with a single Gaussian function and plot the peak center as shown in Fig. 3d. Subsequently, we apply equation (1) to fit the data within the temperature intervals of [11, 300]K and [65, 300]K, corresponding to the black and blue lines in Fig. 3d, respectively (see SI). The outcomes of these fits are different by only a few hundredths of a percent, indicating that the magnon-phonon coupling in this mode is not pronounced. On the other hand, as temperature increases, the sharp peaks are broadened to a single broad peak(fig.3c), meaning the phonon-phonon coupling is destroying the coherence of the spikes, and is consistent with previous reports at room temperature where only a single broad peak is observed \cite{Tian2016, Sun2018}. The temperature dependence of separate comb spikes is also fitted using Gaussian functions, showing a slow decay of energy with a small broadening of the peak width that is consistent with the 3-phonon scattering model(see SI). This result indicates that the phonon-phonon coupling will destroy the frequency comb by broadening the individual peaks.

\begin{figure}%[t]
\centering
\includegraphics[scale=.5]{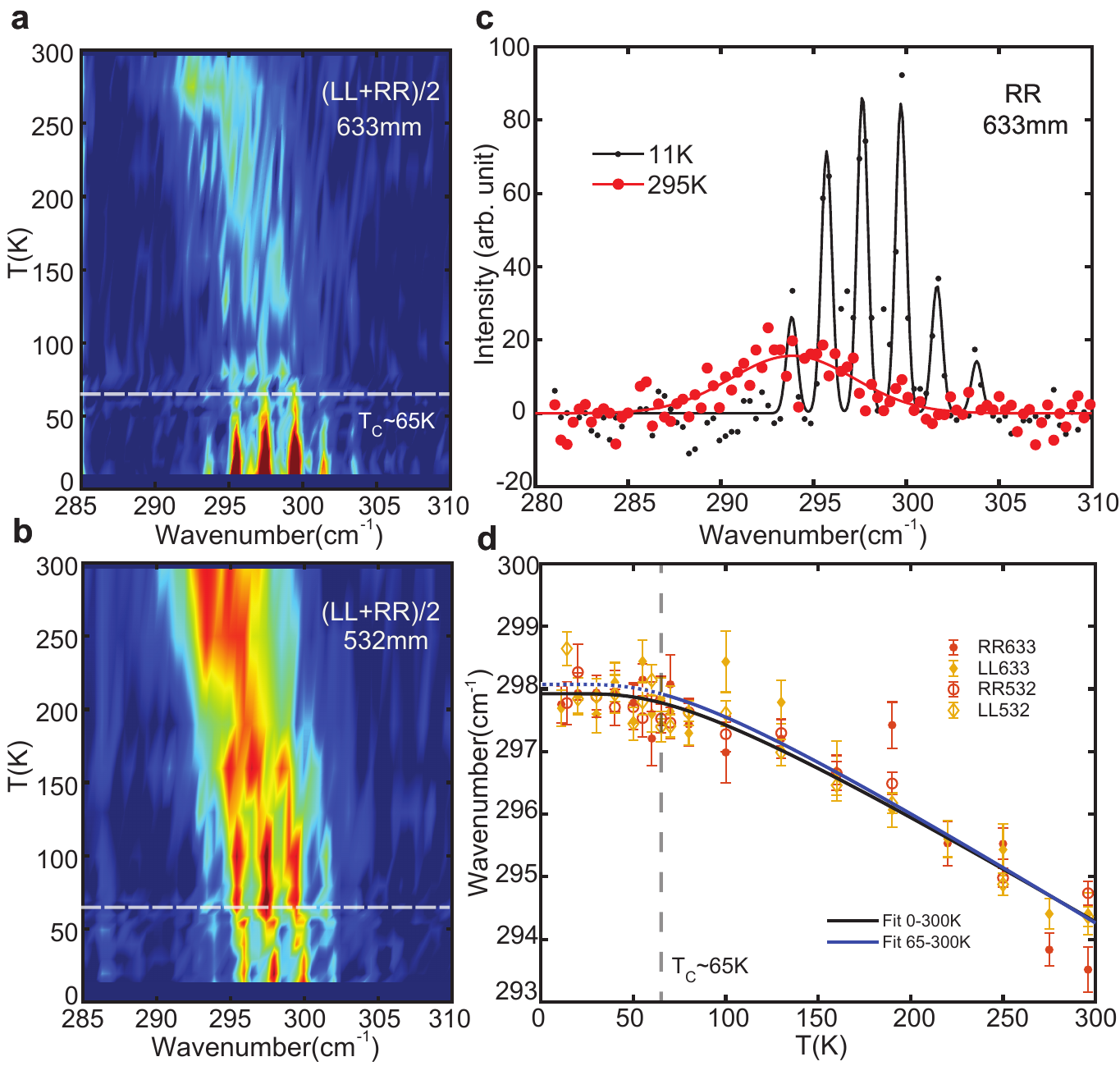}
\caption{\label{fig3} \textbf{Temperature dependence of phonon frequency combs in CrGeTe$_3$.} (a, b) Temperature dependence of the $A_g^5$ mode, with 633nm and 532nm lasers respectively. (c) A comparison of the $A_g^5$ mode between 11K and 295K. (d) Fitting of the temperature dependence using eqn. (1) in the main text. The black solid line shows the fitting with the whole dataset, while the blue line shows the fitting with only $T>$ 65K data.}
\end{figure}

\begin{figure*}
\includegraphics[scale=.08]{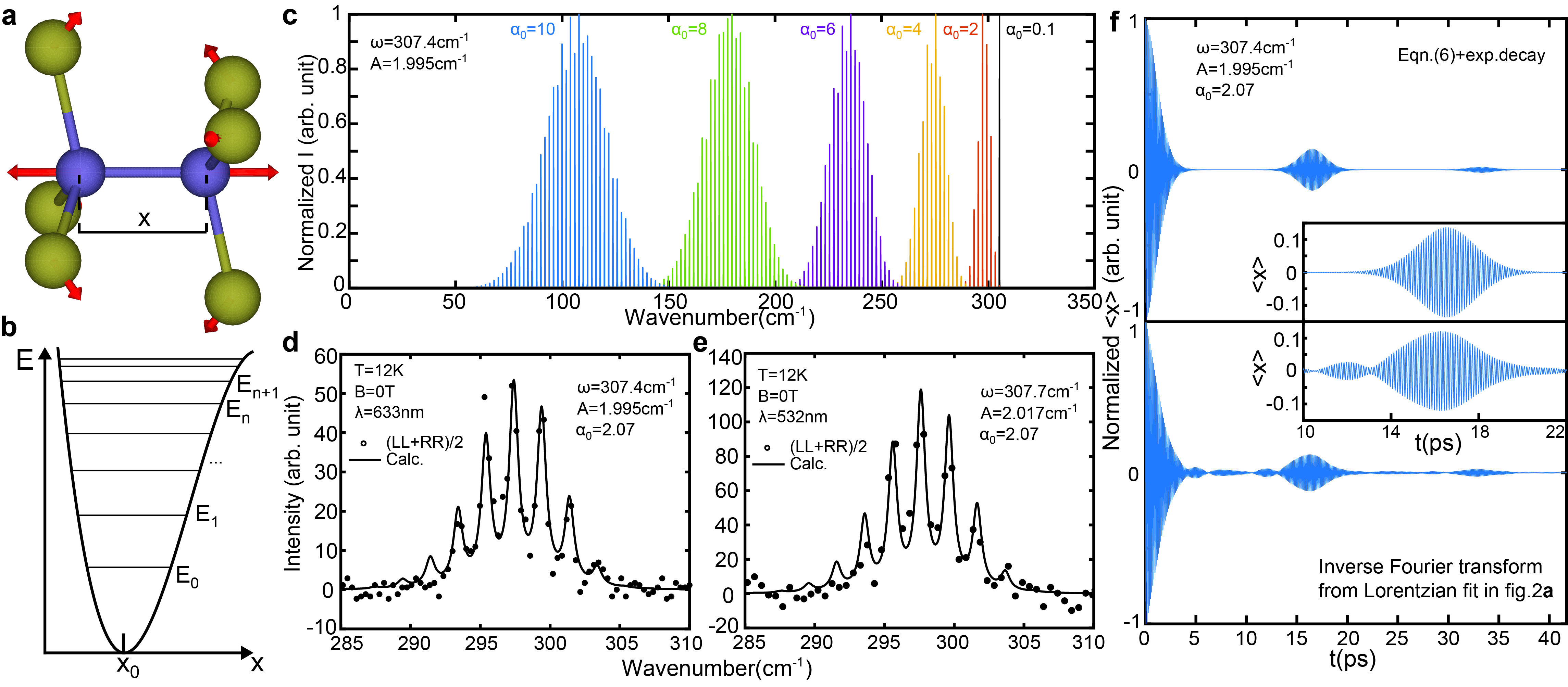}
\caption{\label{fig1} \textbf{Quantum model of phonon frequency combs.} (a) An illustration of the detailed oscillation pattern of the $A_g^5$ mode on the Ge$_2$Te$_6$ cluster. The arrows are not proportional to the actual atomic oscillation amplitude. (b) Schematics of the anharmonic potential and associated energy levels. (c) Calculated frequency comb Raman spectra from eqn. (6) with different $\alpha_0$ values. (d, e) Fitting of the frequency comb with the Fourier transform of eqn. (6) for the 633nm and 532nm data, respectively. (f) Simulation of the atomic movement in the real space for the frequency comb in (d), compared with inverse Fourier transform from the fit in fig.2a. The inset highlights the first revival of the oscillation.}
\end{figure*}

\section*{A quantum model for phonon frequency combs}\label{sec2}

To understand the occurrence of the phonon frequency comb specifically at the $A_g^5$ mode of CrGeTe$_3$ and not in other modes or materials, we analyze the properties of this phonon mode carefully. From previous $ab$ $initio$ calculations on the eigenmodes, this mode is characterized by the head-to-head motion of Ge dimers located at the center of the Cr hexagons, accompanied by the movement of Te atoms, while the Cr atoms surrounding these movements remain mostly stationary (fig.4a)\cite{Milosavljević2018}; Additionally, it can be revealed from the phonon energy calculations that this phonon branch has a very narrow, almost flat dispersion \cite{Zhang2019, Chen2022}. This indicates that the $A_g^5$ phonon mode is extremely localized, i.e., the oscillation of one Ge$_2$Te$_6$ cluster has minimal correlation effect on another. Since one phonon mode corresponds to only one degree of freedom, we can simplify this phonon mode into a collection of isolated quantum oscillators characterized by one variable $x$ which is the Ge-Ge distance relative to its equilibrium value (fig.4a). From the fact that typical optical and phonon frequency comb generation requires nonlinearity, we assume a small, cubic nonlinear term in the harmonic oscillator Hamiltonian:
\begin{equation}
H=\frac{p^2}{2\mu}+\frac{1}{2}\mu\omega^2x^2+\lambda x^3,
\end{equation}
where $p$ is the momentum of the mode, $\mu$ is an effective mass of the oscillator, and $\lambda$ is the cubic anharmonicity. Using second-order perturbation theory, we can calculate the energy difference between adjacent eigenstates $|n\rangle$ and $|n+1\rangle$ as
\begin{equation}
E_{n+1}-E_n = \hbar\omega-\hbar A(n+1),
\end{equation}      
where $A$ is positive and is proportional to $\lambda^2$(fig.4b). The subsequent step to generate frequency combs is to achieve a proper superposition of these states. To do this, we apply a semi-classical approach to the Raman scattering process. Specifically, we treat the oscillations within the phonon mode as purely quantum mechanical, whereas the interaction between atomic oscillations and the Raman laser is regarded as classical. A suitable start for the analysis is to assume the Raman laser can excite a coherent state of the oscillator, which is a classical analog in quantum oscillators \cite{Kuznetsov1994,Tutunnikov2021}: 
\begin{equation}
 |\alpha_0\rangle=e^{-\frac{|\alpha_0|^2}{2}}\sum_{n=0}^{\infty}\frac{\alpha_0^n}{\sqrt{n!}}|n\rangle
\end{equation}
where $\alpha_0$ is a complex eigenvalue for the coherent state, whose absolute value $|\alpha_0|$ is analogous to the amplitude for the coherent state. In the oscillation process, every eigenstate $|n\rangle$ in the coherent state receives a phase factor $e^{-iE_nt/\hbar}$, and the time evolution of the expectation values can be calculated with the coherent state. Here we are interested in the expectation value of $x$ as the classical response for Raman scattering:
\begin{equation}
\langle x \rangle_t=\langle a^\dagger+a \rangle_t=e^{-|\alpha_0|^2}\sum_{n=0}^{\infty}\frac{|\alpha_0|^{2n}\alpha_0}{n!}e^{-i(E_{n+1}-E_n)t/\hbar}+c.c.
\end{equation}
where $a^\dagger$ and $a$ are the creation and annihilation operators respectively, and $c.c.$ stands for 'complex conjugate'. Putting $E_{n+1}-E_n=\hbar\omega-\hbar A(n+1)$ into the equation, we get 
\begin{equation}
\langle x \rangle_t=\alpha_0 \exp[-i(\omega-A)t+|\alpha_0|^2(e^{iAt}-1)]+c.c.
\end{equation}
Equation (6) describes the 'classical' motion of atoms within the phonon mode, and a sample plot of this equation is illustrated in Figure 4(f). The Fourier transformation of Equation (6), as shown in Figure 4(c), transforms the classical atomic movement of the phonon mode into its frequency domain, enabling a detailed analysis of the spectrum. In this spectrum, the parameter $\omega$ indicates the bare frequency of the oscillation, and the parameter $A$ is equal to the distance between adjacent spikes in the frequency comb. Meanwhile, $\alpha_0$ acts as a 'window' that determines which part of the frequency comb is selected to appear in the spectrum. For small $\alpha_0$, the spectral weight is mainly located on the frequency $\omega-A$ which is the upper limit of the comb frequency. As $\alpha_0$ increases, the spectrum goes through a widening and a decrease in energy, indicating the nonlinear terms taking effect. 

\begin{figure*}
\includegraphics[scale=.12]{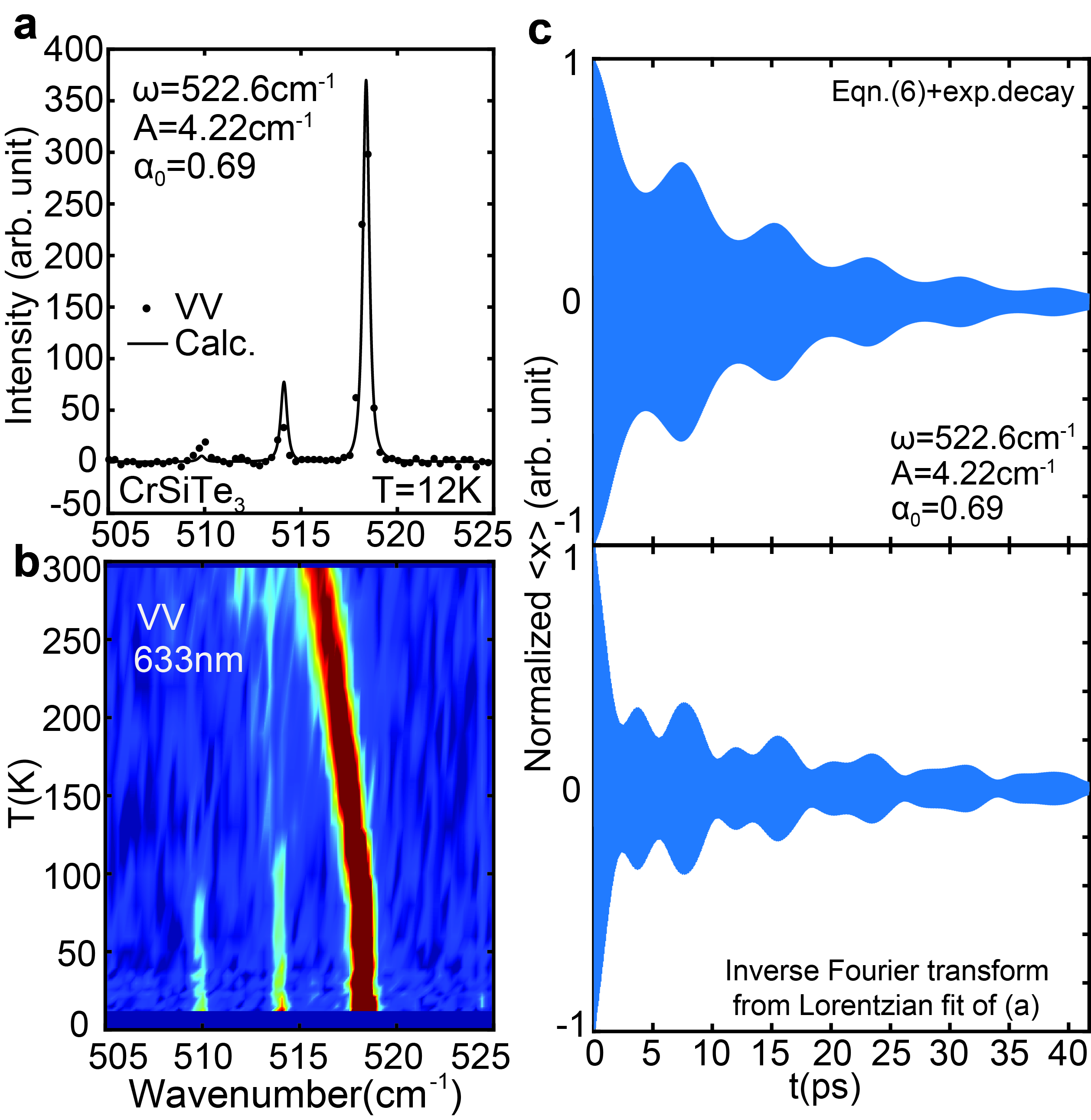}
\caption{\label{fig1} \textbf{Phonon frequency combs in CrSiTe$_3$.}(a) Raman spectrum of the $A_g^5$ mode in CrSiTe$_3$. The black solid line is a fit with the Fourier transform of eqn.(6). (b) Temperature dependence of the $A_g^5$ mode. (c)Simulation of the atomic movement for both theoretical and observed phonon frequency combs.}
\end{figure*}

In terms of the Raman intensity, if we assume that the amplitude of the scattered Raman photons are proportional to $\langle x \rangle_t$, then the Raman intensity will be proportional to $(\langle x \rangle_\omega)^2$, where $\langle x \rangle_\omega$ is the Fourier transform of $\langle x \rangle_t$ into the frequency space. To fit the observed spectrum in CrGeTe$_3$, we utilize this model with the inter-spike distance $A$=1.995cm$^{-1}$ determined from prior analysis. From the simulation shown in Fig. 4(c), the upper limit of the frequency comb is $\omega-A$, which may or may not be the $\#$1 mode denoted in Fig. 2a. Therefore, we adjust both $\alpha_0$ and $\omega$ and apply the Fourier transformation of Equation (6) to simulate the spectrum. We fit the theoretical predictions with the observed data by convolving the spectrum with a Lorentzian. The optimal fitting parameters and the comparison between the calculated and experimental spectra are presented in Figures 4d and 4e for the 633nm and 532nm scattering data respectively, showing that our model can well reproduce the observed frequency comb features. The fitting parameters suggest that the $\#1$ mode is one spike below the upper limit of $\omega-A$, which is equal to the $E_0$ in the linearity fit shown in Fig. 2d. Furthermore, to compare the real space oscillation pattern between the theory and experiment, we calculate the inverse Fourier transform of the Lorentzian fits in fig.2a, and plot it with the calculation directly from eqn. (6) as shown in Fig. 4(f). These two patterns closely resemble each other, suggesting that our model correctly captures the temporal evolution of the oscillations. Notably, the small linewidth of the observed frequency combs allows the signals to revive after more than 300 oscillations. When accounting for instrumental resolution, the lifetime of these oscillations extends to over 500 periods (see SI). This indicates that the anharmonicity parameter $A$ contributes minimally to phonon decay and primarily drives the parametric oscillations.

\section*{Frequency combs in CrSiTe$_3$}\label{sec3}

With the theory in mind, we conducted the same Raman scattering experiment on CrSiTe$_3$ which shares the same structure as CrGeTe$_3$, only with the Ge atoms replaced by Si\cite{Ouvrard1988}. This replacement has two effects: (a) it changes the atomic potential as well as anharmonicity, and (b) it alters the electron-phonon coupling in the $A_g^5$ mode, giving rise to a different driving amplitude. Figure 5 summarizes the Raman scattering results on the $A_g^5$ mode in CrSiTe$_3$. Similar to CrGeTe$_3$, the $A_g^5$ mode in CrSiTe$_3$ has a flat dispersion at $\sim$500cm$^{-1}$\cite{Casto2015}, which is consistent with our observation of the main phonon peak at 518.4cm$^{-1}$(Fig.5a). Notably, two satellite peaks with weaker intensity are observed at 514.2cm$^{-1}$ and 509.9cm$^{-1}$, aligning with the model’s prediction, as the frequency comb feature occurs only below the bare phonon frequency. A further simulation of the data with a smaller $\alpha_0$, and a larger $A$ can well reproduce the spectrum(Fig.5a), indicating that the model from eqn.(6) has the ability to describe the frequency comb in both compounds, only with different parameters. Temperature dependence of the excitation is performed as well, revealing the survival of frequency comb up to 100K (Fig.5b). The temporal evolution of the frequency combs in CrSiTe$_3$ features a weaker revival of the signal, consistent with the fact that smaller $\alpha_0$ represents less reveal of anharmonicity. Nevertheless, the signal still survives long time as reflected by the small linewidth observed in the Raman spectrum(Fig.5c).

%A noteworthy observation from the theory is that the driving frequency and amplitude do not alter the comb distance, but only the anharmonicity $A$ does. This is different from the behavior typically seen in microresonator-generated 'classical' phonon frequency combs, where the frequency and amplitude of the driving force can significantly affect the comb distance and other characteristics\cite{Ganesan2017,Qi2020}. This is also consistent with the fact that two different incident Raman photon wavelengths yield the same comb distance within the fitting error, only with different Raman intensities. This manifests that our observation of frequency combs is from purely internal quantum origins and is independent of the external driving forces or probes. 

\section*{Conclusion and outlook}\label{sec3}

In this study, we have demonstrated the spontaneous formation of phonon frequency combs in the van der Waals materials CrGeTe$_3$ and CrSiTe$_3$ using high-resolution Raman scattering. Our simplified analysis, which includes anharmonic interactions within the atomic potential between Si/Ge dimers, matches the experimental results well. This suggests that the anharmonicity in these systems is not strongly linked to rapid phonon decay, a finding that raises interesting theoretical questions. Further analysis, particularly involving detailed phonon eigenvector calculations, will be crucial to fully resolve the mechanisms behind this reduced decay and parametric oscillations.

Experimentally, we have shown that frequency-based technologies such as Raman scattering and infrared spectroscopy serves as a powerful probe for investigating high-frequency nonlinear phonon modes, particularly in the 5-15THz range, where temporal techniques such as ultrafast pulsed lasers often face limitations in pulse bandwidth. Specifically, our work highlights Raman scattering as an effective tool for studying the dynamics of these modes and provides a novel way to explore frequency comb generation in optical phonon branches of crystalline solids, with implicit phonon coherence. To confirm this coherence, the development of new technologies that enable ultrafast phonon excitation in the 5–15 THz range will be necessary\cite{Xu2023}, which is beyond the scope of this work. These findings offer new insight into the generation and control of long-lived, tunable nonlinear phonons, with potential applications in ultrafast material control and phonon-based technologies.

\section*{Methods}%\label{sec11}

Single crystals of CrGeTe$_3$ were grown using a self-flux method. % [1] What is this?
Cr, Ge, and Te powders in a mass ratio of 1:2:20 were mixed and sealed in a quartz tube under vacuum. The mixture was heated to 1050°C and then cooled down to 475°C. The crystals were harvested after removing the Ge-Te flux. CrGeTe$_3$ crystals were mounted in a closed-cycle helium cryostat (from Cryo Industries of America, Inc.) with a window for optical access. Surface layers of the crystals were exfoliated using adhesive tapes to obtain fresh sample surfaces. Single crystals of CrSiTe$_3$ were grown using a self-flux method described in ref. \cite{Casto2015}. Raman measurements were conducted using both 532 nm and 633 nm lasers with 0.3 mW power. The laser beam was focused to a spot size of 2-3 $\mu$m on the probed sample using a 40x objective
lens. Raman spectra were acquired using a Horiba LabRAM HR Evolution Raman microscope that is equipped with an 1800 grooves/mm grating (with instrumental resolution of 0.4 cm$^{-1}$) and a thermoelectric-cooled CCD. A superconducting magnet (also from Cryo Industries of America, Inc.) was interfaced with the variable-temperature helium cryostat to apply an out-of-plane magnetic field. All measurements were performed at base pressure lower than $7\times10^{-7}$ torr.

\section*{Acknowledgements}

L.C. and R.H. are grateful to Liuyan Zhao for fruitful discussions. The work at the University of California, Berkeley and Lawrence Berkeley National Laboratory was supported by the U.S. DOE under contract no. DE-AC02-05-CH11231 within the Quantum Materials Program (KC2202) (R.J.B.). G.Y., C.N., and R.H. are supported by NSF Grants No. DMR-2104036 and No. DMR-2300640. X. C. Pan and G. Cheng would like to thank KAKENHI 22H00278 and 24K16998. D.M. and J.Q.Y acknowledge  support  from  the  US  Department  of Energy, Office of Science, Basic Energy Sciences, Materials Science and Engineering Division. G.C., Y.P.C. and D.M. also acknowledge partial support from Univ. of Southern California via Multidisciplinary University Research Initiatives (MURI) Program (Award No. FA9550-20-1-0322).

\section*{Supplemental Information}

\subsection*{I. Calculation of anharmonic oscillators}

The Hamiltonian for anharmonic oscillators is

\begin{equation}
    H=\frac{p^2}{2\mu}+\frac{1}{2}\mu\omega^2x^2+\lambda x^3,
\end{equation}
and we denote the harmonic eigenstates as $|n\rangle$. It is easy to see that the first-order perturbation energy for every eigenstate $\langle n|\lambda x^3|n\rangle$ is zero since the cubic term contains an odd number of operators. Using second-order perturbation theory, we can get the modified energy levels
\begin{equation}
\begin{split}
    E_n = \hbar\omega(n+\frac{1}{2})-\frac{1}{8}\frac{\lambda^2\hbar^2}{\mu^3\omega^4}[30(n+\frac{1}{2})^2+\frac{7}{2}]
\end{split}
\end{equation}
with
\begin{equation}
\begin{split}
    E_{n+1}-E_n = \hbar\omega-\frac{15}{2}\frac{\lambda^2\hbar^2}{\mu^3\omega^4}(n+1)
\end{split}
\end{equation}
Denoting $A= \frac{15}{2}\frac{\lambda^2\hbar}{\mu^3\omega^4}$, we get
\begin{equation}
\begin{split}
    E_{n+1}-E_n = \hbar\omega-\hbar A(n+1),
\end{split}
\end{equation}
and $A\geq0$. 

We then calculate the effect of this energy modulation on the time evolution of $x$ in the coherent state. The coherent state is written as
\begin{equation}
\begin{split}
    |\alpha_0\rangle=e^{-\frac{|\alpha_0|^2}{2}}\sum_{n=0}^{\infty}\frac{\alpha_0^n}{\sqrt{n!}}|n\rangle
\end{split}
\end{equation}
and its time evolution is
\begin{equation}
\begin{split}
    |\alpha_0\rangle_t=e^{-\frac{|\alpha_0|^2}{2}}\sum_{n=0}^{\infty}\frac{\alpha_0^n}{\sqrt{n!}}e^{-iE_nt/\hbar}|n\rangle.
\end{split}
\end{equation}
Using this, we can calculate the time evolution of the annihilation operator $\langle a \rangle$:
\begin{equation}
\begin{split}
    \langle a \rangle_t =& \langle \alpha_0^* | a | \alpha_0 \rangle_t\\
    =&e^{-|\alpha_0|^2}\sum_{k=0}^{\infty}\sum_{n=1}^{\infty}\frac{\alpha_0^{*k}}{\sqrt{k!}}\frac{\alpha_0^n}{\sqrt{(n-1)!}}e^{-i(E_k-E_n)t/\hbar}\langle k|n-1\rangle\\
    =&e^{-|\alpha_0|^2}\sum_{n=0}^{\infty}\frac{{|\alpha_0|^{2n}}\alpha_0^*}{n!}e^{i(E_{n+1}-E_n)t/\hbar}\\
    =&e^{-|\alpha_0|^2}\sum_{n=0}^{\infty}\frac{{|\alpha_0|^{2n}}\alpha_0}{n!}e^{-i(\omega-A(n+1))t}\\
    =&\alpha_0 e^{-|\alpha_0|^2-i(\omega-A)t}\sum_{n=0}^{\infty}\frac{{|\alpha_0|^{2n}}}{n!}e^{inAt},\\
\end{split}
\end{equation}
where we used $\langle k|n-1\rangle = \delta(k,n-1)$ and $E_{n+1}-E_n = \hbar\omega-\hbar A(n+1)$. Notice that $\sum_{n=0}^{\infty}\frac{{|\alpha_0|^{2n}}}{n!}e^{inAt}$ is a Taylor series expansion of $\exp(|\alpha_0|^2e^{iAt})$, therefore we get

\begin{equation}
\begin{split}
    \langle a \rangle_t =&\alpha_0 e^{-|\alpha_0|^2-i(\omega-A)t}\exp(|\alpha_0|^2e^{iAt})\\
    =&\alpha_0 \exp[-i(\omega-A)t+|\alpha_0|^2(e^{iAt}-1)],
\end{split}
\end{equation}

which leads to the equation (6) in the main text. A few sanity checks can be made here, for example, for $A=0$, the equation changes to the ordinary $e^{-i\omega t}$ oscillation, and the change of $\alpha_0$ will not make a difference in the oscillation frequency. For nonzero $A$, at small $\alpha_0$ the $|\alpha_0|^2(e^{iAt}-1)$ term is less pronounced, indicating that the nonlinear effect is only visible at larger oscillation amplitude, which is also consistent with one's impression on nonlinear optics.

\subsection*{II. Raman tensor and selection rules of CrGeTe$_3$}

Without magnetism, the space group of CrGeTe$_3$ gives 10 Raman active modes, including five $E_g$ modes and five $A_g$ modes. The $E_g$ modes are doubly degenerate and can be further split into $E_{1g}$ and $E_{2g}$ modes. The Raman tensor of these modes can be written as\cite{Tian2016}:
\begin{equation}
A_g=\begin{bmatrix}
a & 0 & 0\\
0 & a & 0\\
0 & 0 & b
\end{bmatrix},
E_{1g}=\begin{bmatrix}
c & d & e\\
d & -c & f\\
e & f & 0
\end{bmatrix},
E_{2g}=\begin{bmatrix}
d & -c & -f\\
-c & -d & e\\
-f & e & 0
\end{bmatrix}
\end{equation}
The Raman intensity can then be calculated using the equation

\begin{equation}
I = |\phi^\dagger R \phi|^2
\end{equation}
with $\phi = \frac{1}{\sqrt{2}}\begin{bmatrix}1  \\i \\0 \end{bmatrix}$ and $\frac{1}{\sqrt{2}}\begin{bmatrix}1  \\-i \\0 \end{bmatrix}$ for left-hand and right-hand polarizations, respectively. Replacing the Raman tensor $R$ with the specific Raman tensor shown in eqn. (15), one can get the Raman intensity for each polarization as shown in TABLE. I, which indicates that the $E_g$ and $A_g$ modes are only respectively active in cross-polarization and parallel-polarization channels for circularly polarized lights.

\begin{table}[]
    \centering
    \begin{tabular}{|c|c|c|c|}
    \hline
        Polarization & $A_g$ & $E_{1g}$ & $E_{2g}$ \\
        \hline
        LR & 0 & $c^2+d^2$ & $c^2+d^2$ \\
        LL & $a^2$ &0 & 0 \\
        RL & 0 & $c^2+d^2$ & $c^2+d^2$ \\
        RR & $a^2$ &0 & 0 \\
        \hline
    \end{tabular}
    \caption{Raman intensity calculated from the Raman tensor of CrGeTe$_3$.}
    \label{tab:my_label}
\end{table}

\subsection*{III. Additional data and plots}

\textbf{Detailed field dependence.} Fig.S1 shows the magnetic field dependence of the Raman signals from the 633nm measurements. 

\textbf{Detailed temperature dependence.} Fig.S2 and S3 show the raw data of the temperature dependence of the Raman signals from the 633nm and 532nm measurements, respectively. A single Gaussian peak is used to fit the spectrum to get the 'center of mass' of the intensities, which is used to fit the 3-phonon decay model in Fig.3d in the main text.

\textbf{Temperature dependence of individual peaks.} Several equal-width Lorentzians are arranged to fit the observed frequency combs for temperatures below or equal to 100K. The fit is plotted in Fig. S4 and the fitting parameters, including the peak position and width, are shown in Fig. S5. It can be seen that the peaks are evenly split throughout the temperature range of the fitting. The positions of the peaks fluctuate due to a small uncertainty ($\sim$0.3cm$^{-1}$) in the determination of zero energy, which is also the reason for the $E_0$ difference for the 633nm and 532nm data shown in Fig. 2 of the main text. From the 3-phonon decay theory (eqn.(1)), The fitting parameter is $E(0)$=297.92cm$^{-1}$,$a$=1.917cm$^{-1}$ for the 0$\sim$300K fit, and $E(0)$=298.07cm$^{-1}$,$a$=2.006cm$^{-1}$ for the 65$\sim$300K fit. The FWHM of the peaks in the frequency comb is around 0.8 cm$^{-1}$ at low temperatures, which is almost the same as the FWHM of the low-energy $A_g^1$ mode, indicating that minimal decay is associated with the phonon frequency combs.





{}
\section*{ }

\renewcommand{\figurename}{Figure S\hspace{-3px}}
\setcounter{figure}{0}

\begin{figure}
\includegraphics[scale=0.8]{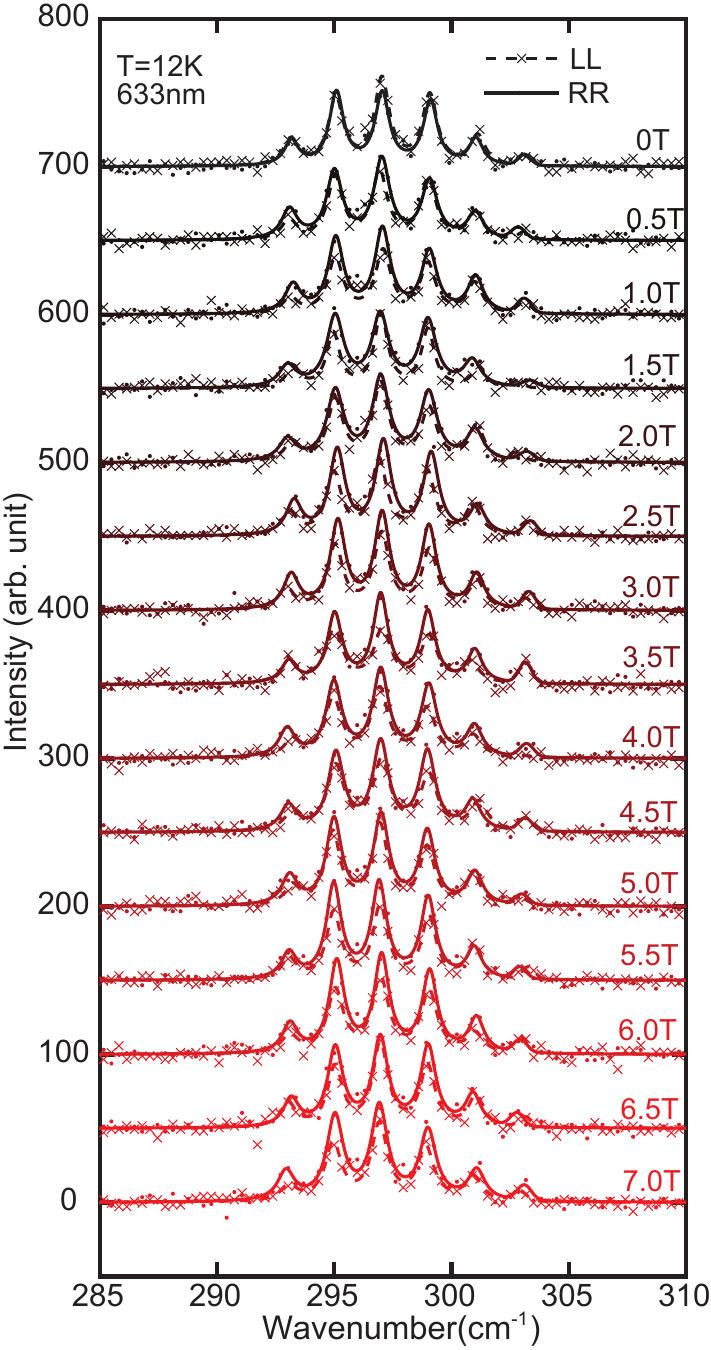}
\caption{Detailed field dependence of the $A_g^5$ phonon mode.}
\end{figure}

\begin{figure}
\includegraphics[scale=.7]{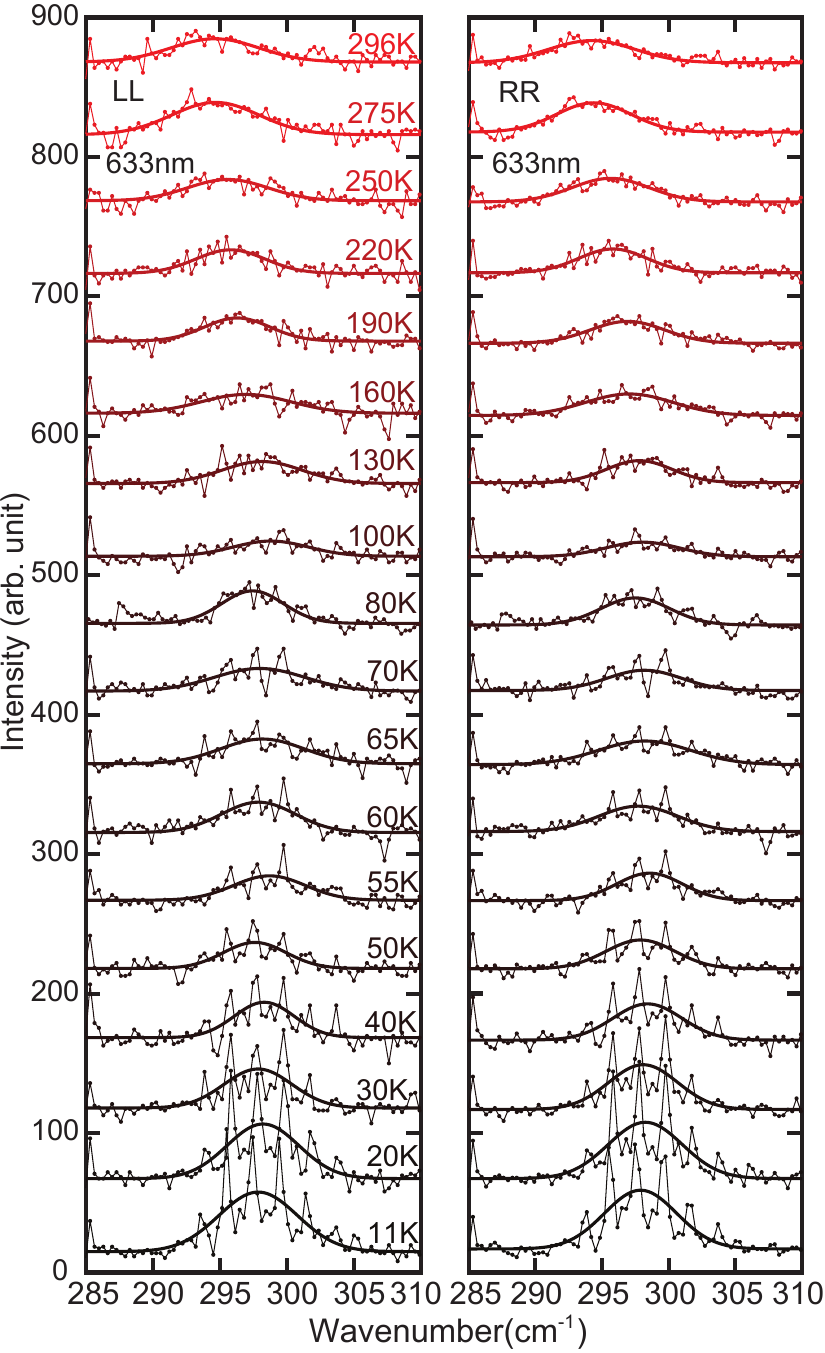}
\caption{Detailed temperature dependence of the $A_g^5$ mode from 633nm-laser experiment. Solid lines are one-peak Gaussian fits.}
\end{figure}

\begin{figure}
\includegraphics[scale=.7]{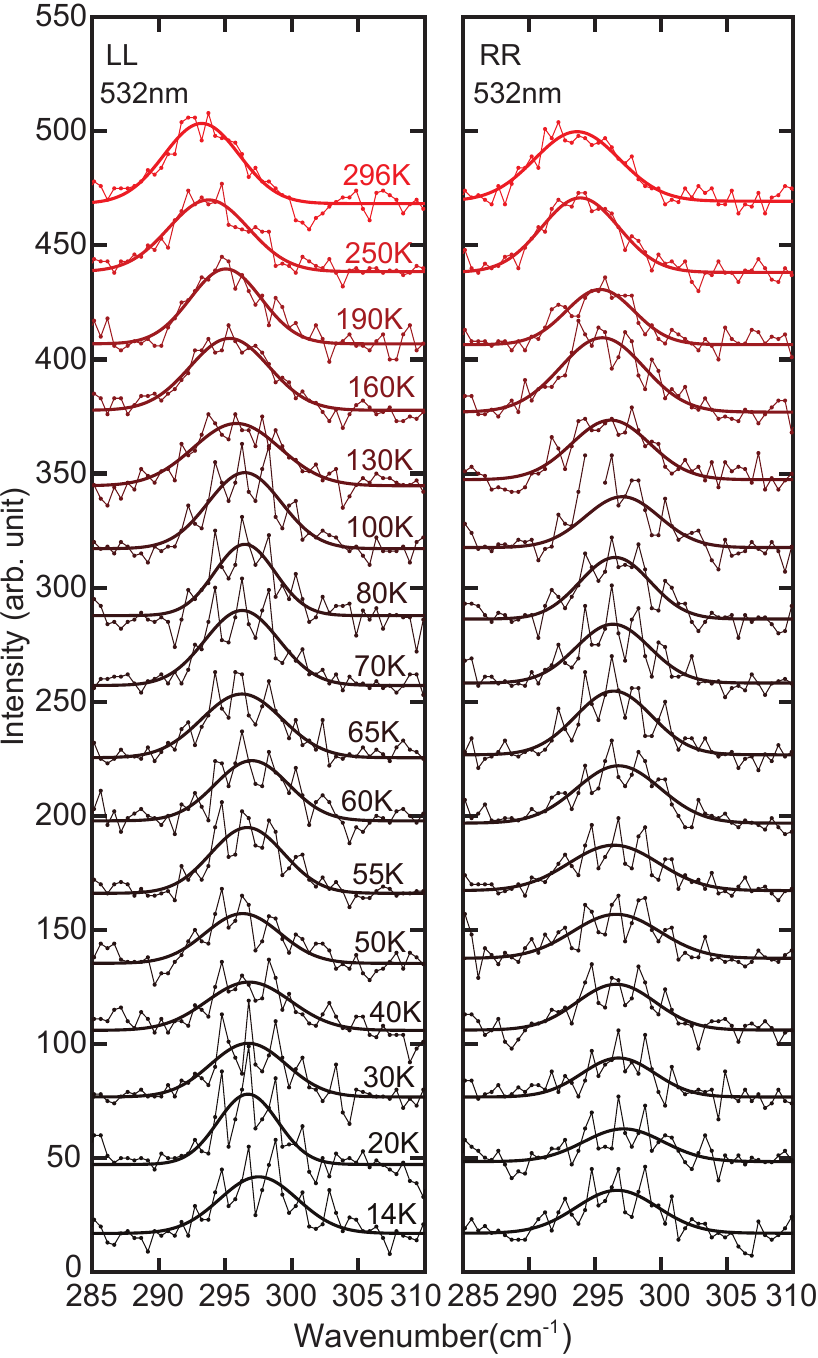}
\caption{Detailed temperature dependence of the $A_g^5$ mode from 532nm-laser experiment. Solid lines are one-peak Gaussian fits. The missing data points at 100K indicate spurious signals whose intensity is beyond the vertical scope of this plot.}
\end{figure}

\begin{figure}
\includegraphics[scale=.15]{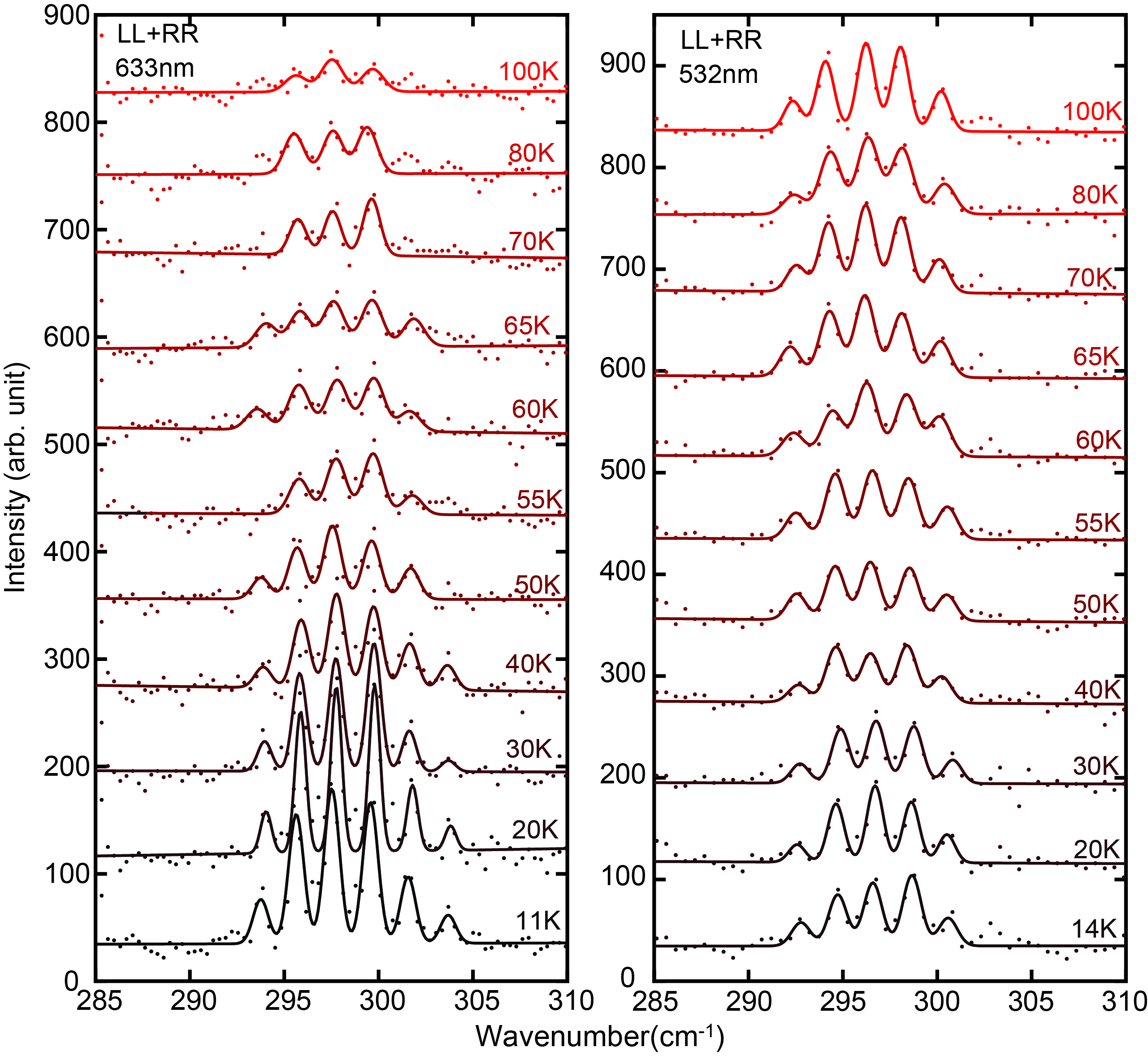}
\caption{Fits on individual frequency comb peaks under temperature $\leq$100K.}
\end{figure}

\begin{figure}
\includegraphics[scale=.12]{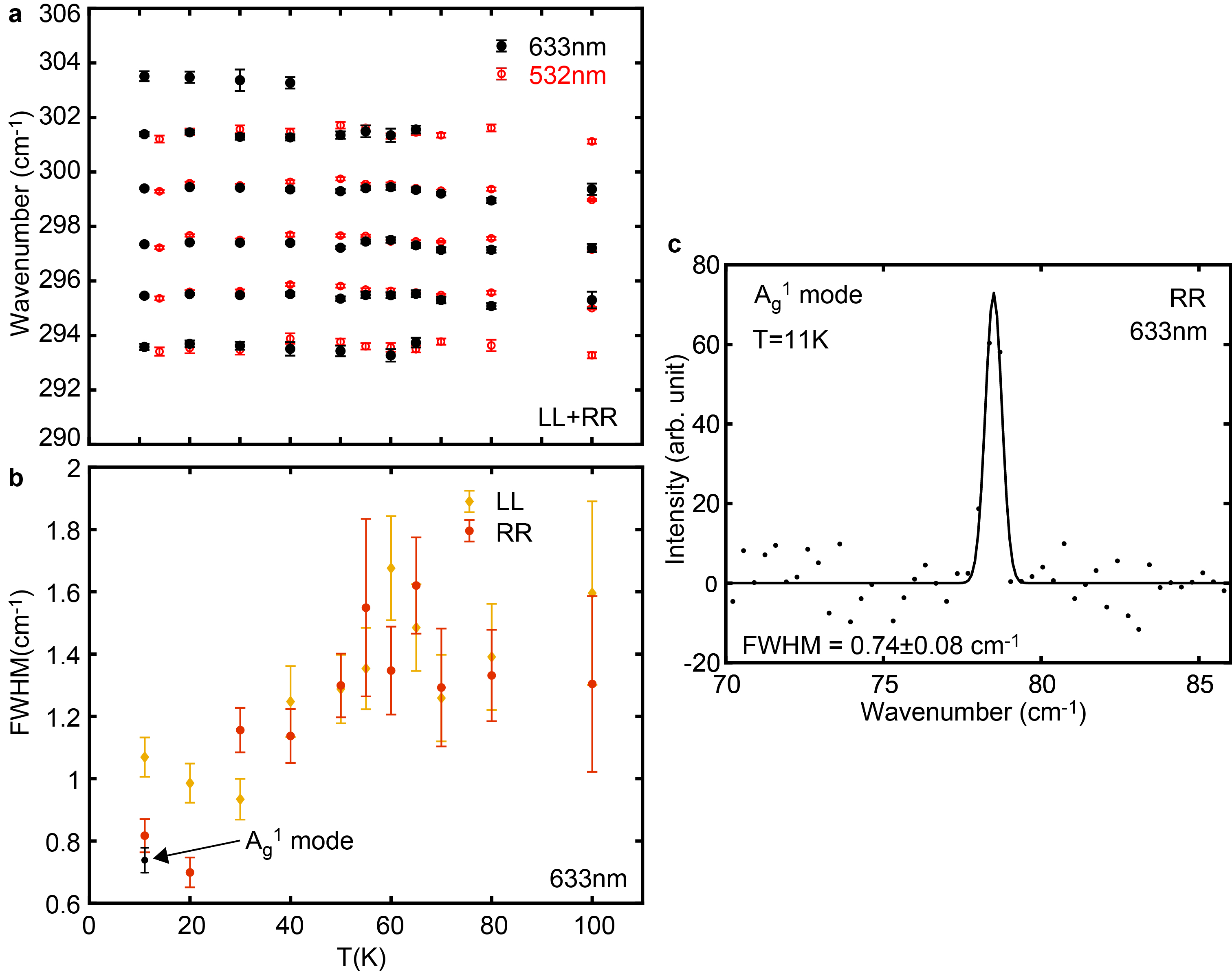}
\caption{(a) The position of the fitted individual peaks in Fig. S4. (b) Temperature dependence of the FWHM for the individual peaks in separate LL/RR channels. The black dot shows the FWHM of the $A_g^1$ mode where minimal phonon decay is expected. (c) Raw data and fitting of the $A_g^1$ mode.}
\end{figure}

\end{document}